\documentclass[a4paper,11pt]{article}
\pdfoutput=1

\usepackage{physics}
\usepackage{empheq}
\usepackage{tensor}
\usepackage{jheppub}
\usepackage{amsmath}
\usepackage{amssymb}
\usepackage{color}
\usepackage{graphicx}

\makeatletter
\gdef\@fpheader{}
\makeatother

\newcommand{\mc}{\mathcal}

\newcommand{\del}{\partial}

\newcommand{\siki}[1]{Eq.~\eqref{#1}}

\newcommand{\wtp}{\widetilde{\phi}}
\newcommand{\hwtp}{\widehat{\widetilde{\phi}}}

\newcommand{\swf}{S^*_{\rm WF}}
\newcommand{\nm}{n_{\rm min}}
\newcommand{\re}[1]{Ref.~#1}

\newcommand{\Sec}[1]{Section \ref{#1}}
\newcommand{\App}[1]{Appendix \ref{#1}}

\begin{document}
\preprint{KUNS-2914, KEK-TH-2386}
\title{Fixed Point Structure of Gradient Flow Exact Renormalization Group for Scalar Field Theories}
\author[a]{Yoshihiko Abe,}
\author[b]{Yu Hamada} 
\author[a]{and Junichi Haruna}

\affiliation[a]{Department of Physics, Kyoto University, Kitashirakawa-Oiwakecho, Kyoto 606-8502, Japan}
\affiliation[b]{KEK Theory Center, High Energy Accelerator Research Organization (KEK), Oho 1-1, Tsukuba, Ibaraki 305-0801, Japan}

\emailAdd{y.abe@gauge.scphys.kyoto-u.ac.jp} 
\emailAdd{yuhamada@post.kek.jp}
\emailAdd{j.haruna@gauge.scphys.kyoto-u.ac.jp}

\abstract{
Gradient Flow Exact Renormalization Group (GFERG) is a framework to define the Wilson action via a gradient flow equation.
We study the fixed point structure of the GFERG equation associated with a general gradient flow equation for scalar field theories and show that it is the same as that of the conventional Wilson-Polchinski (WP) equation in general.
Furthermore, we discuss that the GFERG equation has a similar RG flow structure around a fixed point to the WP equation.
We illustrate these results with the $O(N)$ non-linear sigma model in $4-\epsilon$ dimensions and the Wilson-Fisher fixed point.
}

\date{\today}
\maketitle

\section{Introduction}

Exact Renormalization Group (ERG)~\cite{Wilson:1973jj,Wegner:1972ih,Wilson:1983xri,Polchinski:1983gv,Wetterich:1992yh} (see Refs.~\cite{Berges:2000ew,Polonyi:2001se,Pawlowski:2005xe,Gies:2006wv,Rosten:2010vm} for reviews) is a fundamental and strong framework to study various properties of physical systems using a coarse-graining procedure, namely, varying the energy scale.\footnote{
ERG is also called Functional Renormalization Group or Non-perturbative Renormalization Group.}
It is helpful to investigate critical phenomena and phase structures of various systems in relativistic quantum field theories (QFTs), statistical physics, and condensed matter physics.
In particular, it provides a good understanding of the phase structure of the $O(N)$ linear sigma model~\cite{Tetradis:1993ts,Morris:1999ba,Litim:2002cf,Canet:2002gs,Litim:2010tt,Balog:2019rrg}.
Phase structures of strongly coupled theory such as Quantum Chromodynamics and non-perturbative aspects of QFTs have also been investigated via this framework~\cite{Braun:2005uj,Gies:2005as,Braun:2006jd,Braun:2007bx,Braun:2009gm,Herbst:2010rf,Strodthoff:2011tz,Herbst:2013ail,Mitter:2014wpa,Braun:2014ata,Fu:2019hdw}.
ERG is applied to a field-theoretic approach for quantum gravity, studying non-trivial ultraviolet (UV) fixed points and the continuum limit of gravitational field theories.
The attempt to define quantum gravity based on this approach is called the asymptotic safety scenario~\cite{Weinberg:1976xy,Reuter:1996cp,Souma:1999at} (see, e.g., Ref.~\cite{Eichhorn:2018yfc} for a review).

On the other hand, the gradient flow has recently attracted much attention and is proposed to construct composite operators without the contact divergences due to the coarse-graining via the diffusion~\cite{Luscher:2009eq,Luscher:2010iy,Luscher:2011bx,Luscher:2013cpa}.
It is a one-parameter deformation of fields by a diffusion equation (``gradient flow equation'').
It is shown in Ref.~\cite{Luscher:2011bx} that in the pure Yang-Mills theory, no additional renormalization is needed to keep the correlation functions of the flowed gauge field UV finite in a gauge invariant way once the original theory is renormalized.
The gradient flow method is also applied to matter fermions coupled to the gauge field~\cite{Luscher:2013cpa}, constructing the energy-momentum tensor on the lattice gauge theory~\cite{Suzuki:2013gza,Makino:2014taa}, and supersymmetric models~\cite{Hieda:2017sqq,Kadoh:2018qwg,Kadoh:2019glu}.
In addition to these properties, there are some attempts to interpret the gradient flow as a renormalization group (RG) flow~\cite{Yamamura:2015kva,Aoki:2016ohw,Makino:2018rys,Abe:2018zdc,Sonoda:2019ibh,Carosso:2019qpb,Sonoda:2020vut,Matsumoto:2020lha}.

Recent studies~\cite{Sonoda:2019ibh,Sonoda:2020vut} proposed a new framework to define a Wilson action based on the coarse-graining of the gradient flow, called {\it Gradient Flow Exact Renormalization Group} (GFERG).
In this framework, the Wilson action $S_\tau[\phi]$ at an energy scale $\Lambda = \Lambda_0 e^{-\tau}$, where $\Lambda_0$ is the cutoff scale, is schematically given by the following path integral:
\begin{align}
  e^{S_\tau [\phi]} \sim \int [D\phi']\, \delta (\phi - \varphi [\phi'] ) e^{S_{\tau = 0}[\phi']} \, ,
  \label{e:SchematicDefS}
\end{align}
where $\varphi[\phi']$ is the solution to the gradient flow equation with the initial condition $\eval{\varphi}_{\tau=0}=\phi'$ (details of the definition and calculation are shown in the following sections). 
In GFERG, coarse-graining of the gradient flow is regarded as a ``block spin transformation'' via the diffusion.
A novel characteristics of the GFERG flow is preserving gauge symmetries, even though it has effectively a UV cutoff \cite{Miyakawa:2021hcx,Miyakawa:2021wus}.
From this gauge invariance, GFERG is a promising approach to study gauge theories or quantum gravity non-perturbatively.

Some scalar field theories share similar properties with gauge theories, such as gauge redundancy, asymptotic freedom, or asymptotic safety, and have been studied as their toy models.
A famous example is the $\mathbb{CP}^{N-1}$ model in two-dimensions~\cite{Polyakov:1987hqn}.
This model has $U(1)$ gauge redundancy, asymptotic freedom, and topologically non-trivial configurations like Yang-Mills theories.
Another example is the non-linear sigma model in three dimensions.
This model is renormalizable at the non-perturbative level and has a non-trivial UV completion characterized by the Wilson-Fisher (WF) fixed point~\cite{Wilson:1971dc}.
Studying these scalar field theories within GFERG should give us hints to discuss non-perturbative aspects of gauge theories or quantum gravity.

However, GFERG for general scalar field theories have not been well studied.
This is because the GFERG equation, which is the differential equation for the $\tau$-dependence of the Wilson action $S_\tau$, is defined by the gradient flow equation and takes a different form for each scalar field theory.
The appropriate flow equation heavily depends on details of the theory, such as its symmetry and interactions.
For example, the gradient flow equation for the linear sigma model with the quartic interaction is just given by the diffusion equation~\cite{Capponi:2015ucc},
and the corresponding GFERG equation is shown~\cite{Sonoda:2020vut} to be the Wilson-Polchinski (WP) equation~\cite{Polchinski:1983gv}, which is a well-known ERG equation.
On the other hand, the gradient flow equation for the $O(N)$ non-linear sigma model is proposed as the diffusion equation with extra non-linear terms \cite{Makino:2014sta,Aoki:2014dxa}.
The corresponding GFERG equation is no longer given by the WP equation, which holds not only for this model but also for general scalar field theories.
It is important to study GFERG of these theories in order to get a deeper understanding of them.
This also helps us to understand properties of gauge theories or asymptotically safe gravity.
In particular, fixed points of the GFERG equations and RG flow structure around them are of interest.

In this paper, we introduce the general GFERG equation based on a gradient flow equation of a scalar field theory and study its fixed point structure.
This GFERG equation can be applied to a broad class of non-linear sigma models, whose gradient flow equation is given by a polynomial of the fields, and has  extra non-linear terms compared to the WP equation.
Then, we find that its fixed points appear in the $\tau\to\infty$ limit and are the same as the WP equation.
Moreover, we discuss that the GFERG equation has a similar RG flow structure around a fixed point to the WP equation due to the vanishing of the extra non-linear terms.
We also find that the relevant operators have the same scaling dimensions as those in the WP equation, while the irrelevant operators are not the case.
This result means that the GFERG gives the correct critical exponents and renormalized trajectories.
Therefore, it gives the same predictions as the ordinary ERG for the behavior in the low-energy region.
We illustrate these results with the $O(N)$ non-linear sigma model in $4-\epsilon$ dimensions and the WF fixed point.

This paper is organized as follows.
In \Sec{s:generalGFERG}, we introduce a general gradient flow equation for a scalar field theory and derive the GFERG equation based on this flow equation following Ref.~\cite{Sonoda:2020vut}.
Then, we study the fixed points of the GFERG equation. % in the large flow time limit.
We also investigate the RG flow structure around the fixed points.
In \Sec{s:NLsigma}, we discuss fixed points of the GFERG equation of the non-linear sigma model in $4-\epsilon$ dimensions, as an example.
Then, we illustrate the result in the previous section, focusing on the WF fixed point.
\Sec{s:summary} is devoted to the summary and discussions.
The notation used in this paper is summarized in \App{a:notation}.
%In \App{a:details}, details of the calculation are shown.]
In \App{a:finite-time-action}, we discuss fixed points of the GFERG equation at finite flow time.
We briefly review the RG flow structure around the WF fixed point in the WP equation within the local potential approximation in \App{a:FlowStructureWF}.

\section{GFERG for General Gradient Flow}
\label{s:generalGFERG}
In this section, we discuss GFERG for general gradient flow equation of scalar field theories and show that it has the same fixed point structure as that of the WP equation.
We also compare scaling dimensions of operators between the GFERG equation and the WP equation around the fixed points.
Note that throughout this paper, we work on the dimensionless framework i.e. the dimensionful variables are normalized by the energy scale $\Lambda$.

\subsection{GFERG Equation}
Let us consider a general gradient flow equation defined by the following differential equation:
\begin{align}
\label{e:gGFeq}
    \del_t \varphi_a(t,x) = F_a[\varphi](t,x),\quad \varphi_a(0,x)=\phi_a(x),
\end{align}
where $\phi_a$ is a real scalar field, and $a$ labels all kinds of fields in the theory.
$F_a[\varphi](\tau,x)$ is an arbitrary functional of $\varphi_a$'s.
The variables $t$ and $x$ denote a fictitious time called the flow time and the $D$-dimensional (Euclidean space) coordinate, respectively.
The gradient flow continuously deforms the fields $\phi_a$ defined on the $D$-dimensional Euclidean space along the flow time $t$.

In this paper, we assume that $F_a[\varphi]$ is expanded as a polynomial of $\vec{\varphi}$ like
\begin{multline}
    F_a[\varphi](t,x) = \del_\mu^2 \varphi_a(t,x)
\\
+ \sum_{n=n_{\rm min}}^\infty \int_{x_1,\ldots, x_n} f_a^{a_1,\ldots,a_n}\qty(x;x_1,\ldots,x_n; \del_{x_1},\ldots, \del_{x_n})
    \prod_{j=1}^n \varphi_{a_j}(t,x_j)
\end{multline}
where the expansion coefficient $f_a^{a_1,\ldots,a_n}$ depends on $x_i$'s and contains partial derivatives with respect to them, and $\nm$ is a positive integer larger than one.\footnote{
Note that the linear term is the same as the diffusion equation while the non-linear terms take general forms in the above gradient flow equation. 
In principle, a more general linear term can be considered in the GFERG framework as well. See \re{\cite{Matsumoto:2020lha}} for such cases.
}
For example, gradient flow equations for the non-linear sigma model in two-dimensions are proposed as 
% \begin{equation}
$F_a=\del_\mu^2 \varphi_a-(\varphi_b\del_\mu^2 \varphi_b)\varphi_a$ for $a,b=1,\ldots, N$
% \end{equation}
in Ref.~\cite{Makino:2014sta}, or as
$F_a=\del_\mu^2 \varphi_a + \varphi_a\del_\mu \varphi_b \del_\mu \varphi_b+\flatfrac{\varphi_a(\varphi_b\del_\mu \varphi_b)^2}{(1-(\varphi_c)^2)}$ for $a,b,c=1,\ldots,N-1$
in Ref.~\cite{Aoki:2014dxa}.
In the $O(N)$ linear sigma model, the gradient flow equation is just given by the diffusion equation: $F_a=\del_\mu^2 \varphi_a$ \cite{Capponi:2015ucc}.

\begin{figure}[t]
  \centering
  \includegraphics[width=0.48\textwidth]{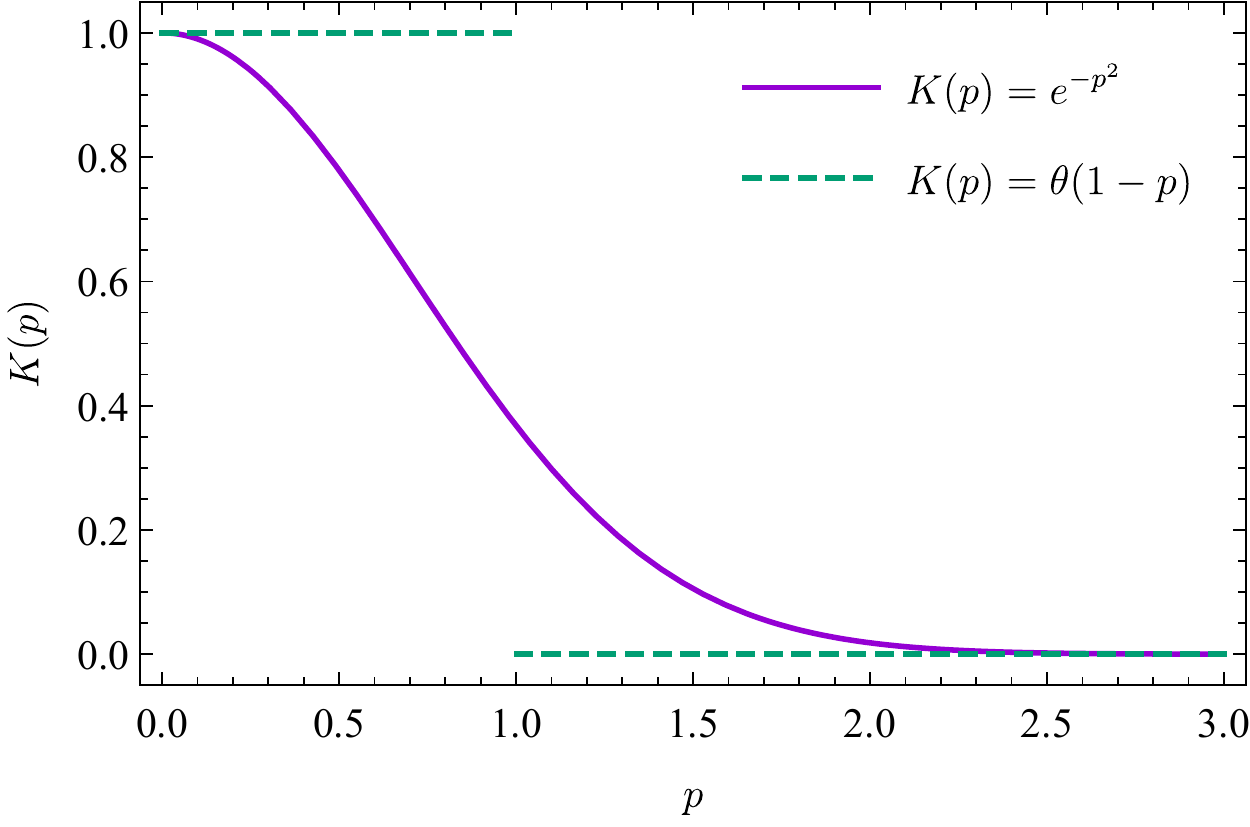}
  \quad
  \includegraphics[width=0.48\textwidth]{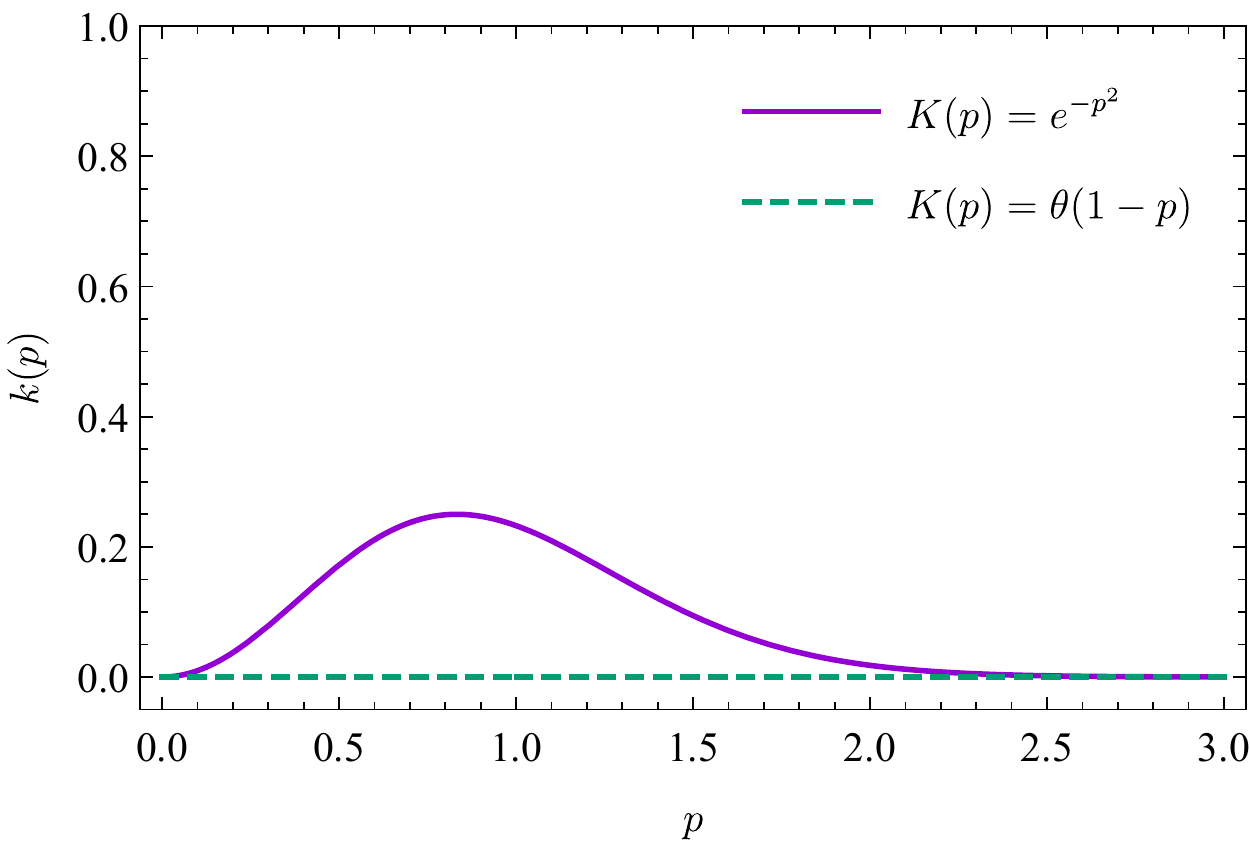}
  \caption{
    Plots of the cutoff functions.
  }
 \label{fig:cutoff}
\end{figure}

Following Ref.~\cite{Sonoda:2020vut}, it is straightforward to define the Wilson action associated with \siki{e:gGFeq} as 
\begin{multline}
\label{e:WilsonAction}
    e^{S_\tau[\phi_a]}=\exp[\int_{x,y} \frac{1}{2} \mc{D}(x-y) \frac{\delta^2}{\delta \phi_a(x)\delta \phi_a(y)}]
    \\
    \times \int [D \phi'_a] \prod_{x',a} \delta(\phi_a(x)-e^{\tau(D-2)/2}Z_\tau^{1/2}\varphi'_a(t,x'e^\tau) )
    \\
    \times\exp[-\int_{x'',y''} \frac{1}{2} \mc{D}(x-y) \frac{\delta^2}{\delta \phi'_a(x'')\delta \phi'_a(y'')}]
    e^{S_{\tau=0}[\phi'_a]} \, ,
\end{multline}
where $\varphi'_a$ is the solution to the general flow equation \siki{e:gGFeq} with the initial condition $\varphi'_a(0,x)=\phi'_a(x)$
and $Z_\tau$ is the wave function renormalization factor depending on $\tau$.
The relation between the flow time $t$ of the gradient flow and $\tau$ in the GFERG equation is given by
\begin{align}
    t\coloneqq e^{2\tau} -1.
\end{align}
$\mathcal{D}(x-y)$ is defined by
\begin{align}
  \mc{D}(x-y) \coloneqq \int_p \ e^{ip(x-y)} \frac{k(p)}{p^2}.
\end{align}
$K(p)$ and $k(p)$ are the cutoff function satisfying
\begin{align}
\label{e:DefKk}
    K(0)=1,\quad K(\infty) = 0,\quad k(0)=0, 
\end{align}
and we set $K(p)=e^{-p^2}$ and $k(p)=K(p)(1-K(p))$ in this paper.
See Fig.~\ref{fig:cutoff} for plots of their $p$-dependence.

Note that \siki{e:WilsonAction} is not invariant under target space diffeomorphism of the fields and seems inapplicable for non-linear sigma models with a curved target space metric.
Instead, this expression should be interpreted as one for non-linear sigma models embedded in higher-dimensional Euclidean space.
This prescription is ensured by Nash's embedding theorem, which states that we can embed an arbitrary Riemann manifold into a Euclidean space $\mathbb{R}^m$ of sufficiently large dimensions $m$ with some constraints for the fields (coordinates of the target space).
Then the consistency under the diffeomorphism does not matter, and the gradient flow equation is required to preserve the constraint instead.

By differentiating $S_\tau$ with respect to $\tau$, we get the GFERG equation for $S_\tau$. 
It is given by
\begin{multline}
\label{e:ERG1}
    \pdv{\tau} e^{S_\tau[\phi_a]}=
    \exp[\lambda(\tau)^2\int_{x,y} \frac{1}{2} \mc{D}(x-y)\frac{\delta^2}{\delta \wtp_a(x)\delta \wtp_a(y)}]
    \\
    \times \int d^Dx' \fdv{\wtp_a(x')}
    \qty[
    -2F_a[\wtp](x)-\qty(\frac{D-2}{2}+ \frac{\eta_\tau}{2}+x'_\rho\del'_\rho)\wtp_a(x')
    ]
    \\
    \times
    \exp[\lambda(\tau)^2\int_{x'',y''} \frac{1}{2} \mc{D}(x''-y'')\frac{\delta^2}{\delta \wtp_a(x'')\delta \wtp_a(y'')}]
    e^{S_{\tau}[\phi'_a]},
\end{multline}
where $\wtp$ is the rescaled field defined as
\begin{align}
    \wtp_a \coloneqq \lambda(\tau) \phi_a
\end{align}
and $\lambda(\tau)$ is given by
\begin{align}
\lambda(\tau) \coloneqq e^{-\tau(D-2)/2}Z_\tau^{-1/2}.
\label{e:DefLambda}
\end{align}
Furthermore, using the relation
\begin{align}
&\hwtp_a(x) \coloneqq \wtp_a(x)+\lambda^2 \int_y \mc{D}(x-y) \fdv{\wtp_a(y)}
\\
&=\exp[\lambda^2\int_{x,y} \frac{1}{2} \mc{D}(x-y)\frac{\delta^2}{\delta \wtp_a(x)\delta \wtp_a(y)}] \wtp_a(x)
  \exp[-\lambda^2\int_{x,y} \frac{1}{2} \mc{D}(x-y)\frac{\delta^2}{\delta \wtp_a(x)\delta \wtp_a(y)}],
\end{align}
\siki{e:ERG1} can be written in a compact form:
\begin{align}
\label{e:CompactGFERG}
\pdv{\tau} e^{S_\tau[\phi_a]}
=\int d^Dx\fdv{\wtp_a(x)}
\qty[
-2F[\hwtp](x) -\qty(\frac{D-2}{2}+ \frac{\eta_\tau}{2}+x_\nu\del_\nu)\hwtp_a(x)
]
e^{S_\tau[\phi_a]},
\end{align}
where the anomalous dimension $\eta_\tau$ is defined as
\begin{align}
  \eta_\tau \coloneqq \frac{d \log Z_\tau}{d\tau}.
\end{align}
More specifically, \siki{e:CompactGFERG} can be written in the following form:
\begin{multline}
\label{e:GFERGeq}
\pdv{t} e^{S_\tau[\phi_a]}
=
\int_p
\biggl\{
    \biggl[
        \qty(2p^2+\frac{D+2}{2}-\frac{\eta_\tau}{2}) \phi_a(p)+p_\mu\pdv{p_\mu} \phi_a(p)
    \biggr]
    \fdv{\phi_a(p)}
    \\
    +\frac{1}{p^2}
    \qty[
%        4p^2k(p)+2p^2 \dv{k(p)}{p^2}- \eta_\tau k(p)
        4p^2k(p)+2p^2 \frac{d k(p)}{d p^2}- \eta_\tau k(p)
    ]
        \frac{1}{2}\frac{\delta^2}{\delta \phi_a(p) \delta \phi_a(-p)}
\biggr\} e^{S_\tau[\phi_a]}
\\
    -2\sum_{n={\nm}}^\infty \lambda(\tau)^{n-1}\int_{x,x_1,\ldots,x_n}
    \fdv{\phi_a(x)}
\biggl\{
    f_a^{a_1,\ldots,a_n}\qty(\phi_{a_1}(x_1)+\int_{y_1} \mc{D}(x_1-y_1) \fdv{\phi_{a_1}(y_1)})\times \ldots
    \\
    \times\qty(\phi_{a_n}(x_n)+\int_{y_n} \mc{D}(x_n-y_n) \fdv{\phi_{a_n}(y_n)})
\biggr\}
e^{S_\tau[\phi_a]}.
\end{multline}
We have ignored ordering of $\delta/\delta\phi_a$ and $\phi_a$ in the first and second lines because it only changes the Wilson action $S_\tau$ by a field-independent constant. 

In Ref.~\cite{Sonoda:2020vut}, the sigma model with a single scalar field is considered, where $F[\varphi]$ is just given by $\del_\mu^2\varphi$.
There, $Z_\tau$ is necessary from the renormalizability of correlation functions of the flowed field $\varphi(t,x)$, that is, to keep the quantity
\begin{align}
    Z_\tau^{n/2}\ev{\exp\qty[-\int_p \frac{k(p)}{p^2}\frac{1}{2}\frac{\delta^2}{\delta \phi(p)\phi(-p)}]\varphi(t,p_1)\cdots \varphi(t,p_n)}_{S_{\tau=0}}
\end{align}
UV-finite after performing the renormalization of the original theory with identification of $t=e^{2\tau}-1$.
The GFERG equation of this model becomes
\begin{multline}
\label{e:WPeq}
\pdv{\tau} e^{S_\tau[\phi_a]}
=
\int_p
\biggl\{
    \biggl[
        \qty(2p^2+\frac{D+2}{2}-\frac{\eta_\tau}{2}) \phi_a(p)+p_\mu\pdv{p_\mu} \phi_a(p)
    \biggr]
    \fdv{\phi_a(p)}
    \\
    +\frac{1}{p^2}
    \qty[
%        4p^2k(p)+2p^2 \dv{k(p)}{p^2}- \eta_\tau k(p)
        4p^2k(p)+2p^2 \frac{d k(p)}{d p^2} - \eta_\tau k(p)
    ]
        \frac{1}{2}\frac{\delta^2}{\delta \phi_a(p) \delta \phi_a(-p)}
\biggr\} e^{S_\tau[\phi_a]},
\end{multline}
which is nothing but the WP equation.
Comparing the general GFERG equation \siki{e:GFERGeq} with this equation, we readily notice that the former has the extra non-linear terms accompanied with one or more factors of $\lambda(\tau)$.
We study their effect on the fixed points and RG flow structure around the fixed points in the following sections.

\subsection{Fixed Points}
\label{s:FP}
In this section, we study fixed points of the GFERG equation \siki{e:GFERGeq} for the general gradient flow equation.
Then, we show that the fixed points of the WP equation \cite{Polchinski:1983gv} appear in the $\tau\to\infty$ limit along the GFERG flow. 

Let us consider the solution $S_\tau$ to the GFERG equation and the limiting value of $S_\tau$ as $\tau\to \infty$.
If $S_\tau$ converges in this limit to some finite action $S^*$, it becomes $\tau$-independent, i.e., $\del_\tau S^*=0$.
Therefore, $S^*$ satisfies
\begin{multline}
0
=
\int_p
\biggl\{
    \biggl[
        \qty(2p^2+\frac{D+2}{2}-\frac{\eta}{2}) \phi_a(p)+p_\mu\pdv{p_\mu} \phi_a(p)
    \biggr]
    \fdv{\phi_a(p)}
    \\
    +\frac{1}{p^2}
    \qty[
        4p^2k(p)+2p^2 \frac{d k(p)}{d p^2} - \eta k(p)
    ]
        \frac{1}{2}\frac{\delta^2}{\delta \phi_a(p) \delta \phi_a(-p)}
\biggr\} e^{S^*}
\\
    -2\sum_{n={\nm}}^\infty \lambda(\infty)^{n-1}\int_{x,x_1,\ldots,x_n}
    \fdv{\phi_a(x)}
\biggl\{
    f_a^{a_1,\ldots,a_n}\qty(\phi_{a_1}(x_1)+\int_{y_1} \mc{D}(x_1-y_1) \fdv{\phi_{a_1}(y_1)})\times \ldots
    \\
    \times\qty(\phi_{a_n}(x_n)+\int_{y_n} \mc{D}(x_n-y_n) \fdv{\phi_{a_n}(y_n)})
\biggr\}
e^{S^*},
\end{multline}
where $\eta$ is the anomalous dimension at the fixed point theory $S^*$ defined as
\begin{align}
    \eta \coloneqq \lim_{\tau\to\infty} \eta_\tau.
\end{align}
We should determine the value of $\lambda(\infty)$ to solve this equation concretely.
Note that the asymptotic behavior of $Z_\tau$ is given by
\begin{align}
    Z_\tau \sim e^{\eta \tau}
\end{align}
as $\tau\to\infty$.
Then the asymptotic behavior of $\lambda(\tau)$ is given by
\begin{align}
    \lambda(\tau) \sim e^{-\tau(D-2+\eta)/2}
\end{align}
from the definition of $\lambda(\tau)$ (\siki{e:DefLambda}).
From this equation, we readily find that the signature of $D-2+\eta$ controls the convergence of $\lambda(\infty)$.
In particular, $\lambda(\infty)$ vanishes when 
\begin{align}
\label{e:ConvCondEta}
    D-2 + \eta > 0.
\end{align}

It is indeed found that $D-2+\eta$ should be positive from physical viewpoints.
Because the fixed point action $S^*$ is invariant under the GFERG flow, it should have the conformal symmetry. 
There, the connected two-point function scales as
\begin{align}
\label{e:TwoPointFunc}
    \ev{\phi(x)\phi(0)}_{\rm conneted} \propto \frac{1}{x^{D-2+\eta}}.
\end{align}
According to the cluster decomposition principle, the two-point function factorizes into the product of one-point functions when $\abs{x}$ goes to infinity.
Therefore we get
\begin{align}
    \ev{\phi(x)\phi(0)}_{\rm conneted} = \ev{\phi(x)\phi(0)} - \ev{\phi(x)}\ev{\phi(0)} \to  0 \quad (\abs{x}\to \infty).
\end{align}
This fact requires that $\eta$ should satisfy \siki{e:ConvCondEta}.

Then, we conclude that $\lambda(\infty)=0$ and $S^*$ satisfies
\begin{multline}
\label{e:FPCondWP}
0
=
\int_p
\biggl\{
    \biggl[
        \qty(2p^2+\frac{D+2}{2}-\frac{\eta_\tau}{2}) \phi_a(p)+p_\mu\pdv{p_\mu} \phi_a(p)
    \biggr]
    \fdv{\phi_a(p)}
    \\
    +\frac{1}{p^2}
    \qty[
        4p^2k(p)+2p^2 \frac{d k(p)}{d p^2} - \eta_\tau k(p)
    ]
        \frac{1}{2}\frac{\delta^2}{\delta \phi_a(p) \delta \phi_a(-p)}
\biggr\} e^{S^*},
\end{multline}    
which is nothing but the fixed point condition of the WP equation. 
Therefore, we find the fixed points of the WP equation appear along the general GFERG flow as $\tau\to\infty$.

Here we give a comment on those fixed points.
Because the GFERG flow depends on the RG flow time, $S_\tau$ cannot stay at $S^*$ at finite flow time.
In other words, even if the Wilson action $S_\tau$ equals to $S^*$ at some finite time $\tau=\tau_1$, $\eval{\del_\tau S_\tau}_{\tau=\tau_1}$ is not zero because of the extra non-linear terms proportional to powers of $\lambda(\tau_1)$ in the GFERG equation.
Therefore, the GFERG equation does not have the same fixed points as those of the WP equation \textit{at a finite flow time}, and they appear \textit{in the large flow time limit}.
Note that the GFERG equation has no fixed point at the finite flow time.
See Sec.~\ref{a:finite-time-action} for a detailed argument.

\subsection{Flow Structure around Fixed Points}
In the previous subsection, we have found that the fixed point action $S^*$ of the WP equation appears in the $\tau\to\infty$ limit along the GFERG flow.
This means that there can be a solution to the GFERG equation that flows into $S^*$ as $\tau\to\infty$.

In this subsection, we study the RG flow structure around a fixed point after a long time.
We investigate the time evolution of the GFERG equation after a long time $\tau = \tau_0\gg 1$ so that $\exp(-\tau_0(D-2+\eta)/2)\ll 1$.
Let us consider perturbing $S_\tau$ from a fixed point of the GFERG equation  at $\tau=\tau_0$ as
\begin{align}
    S_{\tau=\tau_0} = S^* + \sum_{A} \delta c^A \mc{O}_A,
\end{align}
where $\delta c^A$ is a small fluctuation around the fixed point ($\abs{\delta c^A} \ll 1$) and $\mc{O}_A$'s form a complete set of operators (defined later).
If we set $\tau=\tau' + \tau_0$, the GFERG equation is given by
\begin{multline}
\pdv{\tau'} e^{S_{\tau}[\phi_a]}
=
\int_p
\biggl\{
    \biggl[
            \qty(2p^2+\frac{D+2}{2}-\frac{\eta_{\tau}}{2}) \phi_a(p)+p_\mu\pdv{p_\mu} \phi_a(p)
    \biggr]
    \fdv{\phi_a(p)}
    \\
    +\frac{1}{p^2}
    \qty[
        4p^2k(p)+2p^2 \frac{d k(p)}{d p^2} - \eta_\tau k(p)
    ]
        \frac{1}{2}\frac{\delta^2}{\delta \phi_a(p) \delta \phi_a(-p)}
\biggr\} e^{S_\tau[\phi_a]}
\\
    -2\sum_{n={\nm}}^\infty (\lambda(\tau'+\tau_0))^{n-1}\int_{x,x_1,\ldots,x_n}
    \fdv{\phi_a(x)}
\biggl\{
    f_a^{a_1,\ldots,a_n}\qty(\phi_{a_1}(x_1)+\int_{y_1} \mc{D}(x_1-y_1) \fdv{\phi_{a_1}(y_1)})\times \ldots
    \\
    \times\qty(\phi_{a_n}(x_n)+\int_{y_n} \mc{D}(x_n-y_n) \fdv{\phi_{a_n}(y_n)})
\biggr\}
e^{S_\tau[\phi_a]}.
\end{multline}
Note that the asymptotic behavior of $\lambda(\tau'+\tau_0)$ as $\tau_0\to\infty$ is given by  $e^{-\tau'(D-2+\eta)/2}\lambda_0$, where $\lambda_0$ is defined as $\lambda_0\coloneqq \exp(-\tau_0(D-2+\eta)/2)$.
Because both of $\lambda_0$ and $\delta c^A$ are sufficiently small, the solution $S_\tau$ can be expanded in terms of $\lambda_0$ and $\delta c^A$ as 
\begin{align}
    S_{\tau} = S^* + \sum_A (\delta c^A \xi^A(\tau') + \lambda_0^{n_{\rm min}-1} \zeta^A(\tau')) \mc{O}_A + (\text{higher-order terms}).
    \label{e:StLinear}
\end{align}
Note that the leading contribution from $\lambda_0$ in the expansion should be proportional to $\lambda_0^{\nm-1}$ since $\lambda_0$ appears in \siki{e:StLinear} as $\lambda_0^{\nm -1}$ at the leading order.
Substituting this equation into the GFERG equation and focusing on the terms up to the linear order of $\delta c^A$ and $\lambda_0^{\nm -1}$, we get
\begin{multline}
\label{e:StLinear2}
\del_{\tau'}\sum_A (\delta c^A \xi^A(\tau') + \lambda_0^{n_{\rm min}-1} \zeta^A(\tau')) \mc{O}_A 
=\hat{R} \sum_A (\delta c^A \xi^A(\tau') + \lambda_0^{n_{\rm min}-1} \zeta^A(\tau')) \mc{O}_A 
\\
+\lambda_0^{\nm -1} e^{-\tau'(n_{\rm min}-1)(D-2+\eta)/2} H(S^*) , 
\end{multline}
where 
\begin{multline}
    \hat{R}\coloneqq 
\int_p
\biggl\{
    \biggl[
            \qty(2p^2+\frac{D+2}{2}-\frac{\eta}{2}) \phi_a(p)+p_\mu\pdv{p_\mu} \phi_a(p)
    \biggr]
    \fdv{\phi_a(p)}
    \\
    +\frac{1}{p^2}
    \qty[
        4p^2k(p)+2p^2 \frac{d k(p)}{d p^2} - \eta k(p)
    ]
        \fdv{S^*}{\phi_a(p)} \fdv{\phi_a(-p)}
\biggr\}
\end{multline}
and 
\begin{multline}
    H(S^*)  \coloneqq 
-2e^{-S^*}\int_{x,x_1,\ldots,x_{n_{\rm min}}}\fdv{\phi_{a}(x)}
\biggl\{
    f_a^{a_1,\ldots,a_{n_{\rm min}}}\qty(\phi_{a_1}(x_1)+\int_{y_1} \mc{D}(x_1-y_1) \fdv{\phi_{a_1}(y_1)})\times \ldots
    \\
    \times\qty(\phi_{a_{n_{\rm min}}}(x_{n_{\rm min}})+\int_{y_{n_{\rm min}}} \mc{D}(x_{n_{\rm min}}-y_{n_{\rm min}}) \fdv{\phi_{a_n}(y_{n_{\rm min}})})
\biggr\}
e^{S^*}
\end{multline}
Comparing each term of $\order{\lambda_0^{n_{\rm min}-1}}$ and $\order{c^A}$ in the left and right hand sides of \siki{e:StLinear2}, we get
\begin{align}
\label{e:ERGeqTildeS}
\sum_A \delta c^A \del_{\tau'}\xi^A(\tau') \mc{O}_A 
&=\sum_A \delta c^A \xi^A(\tau') \hat{R}  \mc{O}_A 
\\
\sum_A \del_{\tau'}\zeta^A(\tau') \mc{O}_A 
&=\sum_A \zeta^A(\tau') \hat{R} \mc{O}_A 
+e^{-\tau'(n_{\rm min}-1)(D-2+\eta)/2} H(S^*). 
\end{align}
Because $\hat{R}$ is a time-independent operator on the functional space, $\mc{O}_A$ can be taken as its eigenoperator satisfying
\begin{align}
    \hat{R} \mc{O}_A = x_A \mc{O}_A\quad (A=1,2,\ldots)
\end{align}
with its eigenvalue $x_A$.
Note that the index $A$ is not summed in this equation.
Since $\{\mc{O}_A\}$ forms a complete set on the functional space, $H(S^*) $ can be expanded as
\begin{align}
    H(S^*)  & = h^A \mc{O}_A,
    \label{e:DefhA}
\end{align}
where $h^A$ is an expansion coefficient.
Substituting this expression into \siki{e:ERGeqTildeC} and \siki{e:ERGeqC1}, and focusing on the each coefficient of $\mc{O}_A$, we get
\begin{align}
\label{e:ERGeqTildeC}
    \del_{\tau'} \xi^A(\tau') &= x_A \xi^A(\tau'),\\
    \del_{\tau'} \zeta^A(\tau') &= x_A \zeta^A(\tau') + e^{-\tau'(\nm -1)(D-2+\eta)/2} h^A.
\label{e:ERGeqC1}
\end{align}
for each $A=1,2,\cdots$.
The solutions to these equations are given by
\begin{align}
    \xi^A(\tau') &= e^{x_A\tau'},\\
    \zeta^A(\tau') &= \frac{e^{x_A\tau'}-e^{-(n_{\rm min}-1)(D-2+\eta)\tau'/2}}{x_A+(n_{\rm min}-1)(D-2+\eta)/2}h^A,
\end{align}
where we have used the initial conditions $\xi^A(0)=1$ and $\zeta^A (0)=0$.
Finally, we get
\begin{align}
\label{e:SolGFERG}
    S_\tau = S^* 
    + \sum_A \qty(\delta c^A e^{x_A\tau'}+\lambda_0^{n_{\rm min}-1}\frac{e^{x_A\tau'}-e^{-(n_{\rm min}-1)(D-2+\eta)\tau'/2}}{x_A+(n_{\rm min}-1)(D-2+\eta)/2}h^A)\mc{O}_A
\end{align}
to the order of $\lambda_0^{n_{\rm min}-1}$ and $\delta c^A$.
Note that if $x_A+(n_{\rm min}-1)(D-2+\eta)/2=0$, we have
\begin{align}
 \eval{\frac{e^{x_A\tau'}-e^{-(n_{\rm min}-1)(D-2+\eta)\tau'/2}}{x_A+(n_{\rm min}-1)(D-2+\eta)/2}}_{x_A+(n_{\rm min}-1)(D-2+\eta)/2=0}
 =\tau' e^{x_A\tau'}.
\end{align}

Let us discuss scaling dimensions of operators at the fixed point.
In the conventional ERG formalism such as the WP equation, the scaling dimension $d_A$ of an operator $\mc{O}_A$ can be determined from the time-evolution of $S_\tau$ in the direction of $\mc{O}_A$.
For example, let us consider the WP equation, which corresponds to the case $H(S^*)=0$, i.e., $h^A=0$.
There $S_\tau$ is given by
\begin{align}
    S_\tau = S^* + \sum_A \delta c^A e^{x_A\tau'}\mc{O}_A.
\end{align}
Because the time-dependence has a simple form of $e^{x_A\tau'}$, the scaling dimension $d_A$ is defined as $d_A=x_A$.

On the other hand, in the case of the GFERG equation, the time-dependence of $S_\tau$ in $\mc{O}_A$ is a linear combination of $e^{x_A\tau'}$ and $e^{-(n_{\rm min}-1)(D-2+\eta)\tau'/2}$ (see \siki{e:SolGFERG}).
Therefore we should be careful about defining relevant or irrelevant operators and their scaling dimensions in this case.
Recall that whether an operator is relevant or irrelevant corresponds to whether the amplitude of its coupling increases or not (i.e., its linearized flow departs from/converges to the fixed point) as $\tau'$ increases.
Therefore, we find that an operator with positive (negative) $x_A$ should be called relevant (irrelevant) in GFERG like the conventional ERG formalism.

Let us discuss the scaling dimensions of relevant operators from the viewpoint of observable quantities in experiments.
They can be measured by tuning parameters so that the system undergoes a phase transition.
There, the observable quantities are determined by the infrared (IR) behavior of the system, which is described by the renormalized trajectory of the fixed point.
Since the renormalized trajectory is defined by taking the IR limit ($\tau_0\to \infty$) with tuning the relevant (bare) couplings, one is led to consider the $\lambda_0\to 0$ limit to define the scaling dimensions of relevant operators.
Because the time-dependence of $S_t$ in the direction of $\mc{O}_A$ in this limit is the same as the WP equation, we should define their scaling dimensions $d_A$ as $d_A=x_A$ like the conventional ERG formalism.

As for irrelevant operators, their scaling dimensions should be determined as the convergence speed to the fixed point when the theory sits on a critical surface.
From \siki{e:SolGFERG}, we see that for a sufficiently large time $\tau'\gg 1$, the coefficient of the (irrelevant) operator $\mc{O}_A$ is given by
\begin{multline}
\delta c^A e^{x_A\tau'}+\lambda_0^{n_{\rm min}-1}\frac{e^{x_A\tau'}-e^{-(n_{\rm min}-1)(D-2+\eta)\tau'/2}}{x_A+(n_{\rm min}-1)(D-2+\eta)/2}h^A
\\ \propto e^{-\tau'\min(\abs{x_A}, (n_{\rm min}-1)(D-2+\eta)/2)}\quad (\tau'\gg 1).
\end{multline}
Therefore, from the above argument, the scaling dimension $d_A$ of the irrelevant operator $\mc{O}_A$ should be defined as $d_A=-\min(\abs{x_A}, n_{\rm min}(D-2+\eta))$.

Here we comment on the case where the expansion coefficient $h^A$ becomes zero.
In this case, time-dependence of $S_\tau$ in the direction of the corresponding operator is the same as in the WP equation, i.e., $e^{x_A\tau'}\delta c^A$.
Therefore, the scaling dimension of this operator is given by $x_A$ regardless of whether they are relevant or irrelevant.
We will encounter this case in the next section.

\section{Example\ :\ Non-linear Sigma Model in \texorpdfstring{$4-\epsilon$}{4-epsilon} Dimensions}
\label{s:NLsigma}
In this section, we illustrate our results in \Sec{s:generalGFERG} with the $O(N)$ non-linear sigma model in $4-\epsilon$ dimensions and the WF fixed point.

\subsection{GFERG Equation}
Lagrangian of the non-linear sigma model is given by
\begin{align}
    \mathcal{L}=
    \frac{1}{2g^2}\del_\mu \phi_a\del^\mu \phi_a,
\end{align}
where $\phi_a\ (i=1,\ldots,N)$ is a real scalar field constrained by
\begin{align}
    \phi_a\phi_a=1.
\end{align}
Note that this constraint requires the physical degree of freedom to be $N-1$.
$g^2$ is a bare coupling constant.

It is well-known that this model is defined non-perturbatively on the renormalized trajectory of the WF fixed point \cite{Wilson:1971dc}.
The $O(N)$ liner sigma model with the quartic interaction also belongs to this WF universality class after the $O(N)$ symmetry breaks spontaneously down to $O(N-1)$ with a negative mass term.
By setting $D=4-\epsilon$ and solving the fixed point condition \siki{e:FPCondWP} of the WP equation with the $\epsilon$ expansion \cite{Dutta:2020vqo}, we get the action $S^*_{\rm WF}$ at the WF fixed point up to $\order{\epsilon}$ as
\begin{align}
    S^*_{\rm WF} = \int_p \qty( \frac{1}{2}\qty(-\frac{p^2}{K(p)}+m^{2}_*)\phi_a(p)\phi_a(-p) - \frac{\lambda_*}{8} \phi_a(p_1)\phi_a(p_2)\phi_b(p_3)\phi_b(p_4)),
\end{align}
where the (dimensionless) couplings $m_*^2$ and $\lambda_*$ are defined as \footnote{
$\lambda^*$ in \siki{e:ValueML} is different from that in \re{\cite{Dutta:2020vqo}} by a factor $2$.
There seems to be a typo in \re{\cite{Dutta:2020vqo}}. 
}
\begin{align}
    m_*^2 \coloneqq \frac{\epsilon}{4} \frac{N+2}{N+8},
    \quad \lambda_* \coloneqq -\epsilon \frac{8\pi^2}{N+8}.
    \label{e:ValueML}
\end{align}
If the $O(N)$ symmetry is spontaneously broken, the theory contains one massive mode and $N-1$ Nambu-Goldstone (NG) modes.
When we focus on a much lower energy scale compared to the mass of the former, the massive particle becomes sufficiently heavy to decouples from the NG modes.
These remaining NG bosons correspond to the $N-1$ physical degrees of freedom in the non-linear sigma model in the IR region.

The gradient flow equation for this model is given in \re{\cite{Makino:2014sta}} as
\begin{align}
\label{e:GFeqNL}
    \del_t \varphi_a = \del_\mu^2 \varphi_a - (\varphi_b\del_\mu^2\varphi_b) \varphi_a
\end{align}
with the initial condition $\varphi_a(0,x)=\phi_a(x)$ for $a,b=1,\ldots,N$.
An advantage of adopting this flow equation is that in two-dimensions, correlation functions of the flowed field $\varphi_a(t,x)$ are UV-finite {\it without} additional wave function renormalization, i.e., $Z_\tau$ can be set to unity.
On the other hand, $Z_\tau$ cannot be omitted in the present case, which is obvious from the following results in order for the WF fixed point to exist.

The Wilson action of this model can be defined in the same way as \siki{e:WilsonAction} via the solution $\varphi'_i(t,x)$ to \siki{e:GFeqNL} with the initial condition $\varphi'_i(0,x)=\phi_a(x)$.
The GFERG equation associated with this Wilson action is given by
\begin{multline}
\pdv{\tau} e^{S_\tau[\phi_a]} 
=
\int_p
\biggl\{
    \biggl[
        \qty(2p^2+\frac{D+2}{2}-\frac{\eta_\tau}{2})\phi_a(p)+p_\mu\pdv{p_\mu}\phi_a(p)
    \biggr]
    \fdv{\phi_a(p)}
    \\
    +\frac{1}{p^2}
    \qty[
        4p^2k(p)+2p^2 \frac{d k(p)}{d p^2} - \eta_\tau k(p)
    ]
        \frac{1}{2}\frac{\delta^2}{\delta \phi_a(p) \delta \phi_a(-p)}
\biggr\} e^{S_\tau[\phi_a]}
\\
+2 \lambda(\tau)^2 \int_x 
\fdv{\phi_a(x)}
\qty(\phi_b(x) + \int_y \mc{D}(x-y)\fdv{\phi_b(y)})\del_\mu^2\qty(\phi_b(x) + \int_y \mc{D}(x-y)\fdv{\phi_b(y)})
\\
\times 
\qty(\phi_a(x) + \int_y \mc{D}(x-y)\fdv{\phi_a(y)})
e^{S_\tau[\phi_a]}.
\label{e:GFERG_NL}
\end{multline}
The terms in the third and the fourth lines of this equation are peculiar to GFERG, compared to the WP equation.
Note that this GFERG equation is invariant under the global $O(N)$ symmetry and expected to preserve the constraint $\phi_a^2=\text{const.}$ in the correlation functions.

%%%%%%%%%homework
% For an arbitrary $n$, 
% \begin{align}
%     \ll F[\phi] \phi_{a_1}(x_1)\cdots \phi_{a_n}(x_n) \gg_{S_\tau} = 0
% \end{align}
% holds, i.e., $F[\phi]=0$ holds at operator level.
% In the present caes, $F[\phi]= \phi_a(x)^2-1$.

\subsection{Wilson-Fisher Fixed Point}

Let us confirm that the GFERG equation \siki{e:GFERG_NL} has the WF fixed point in the $\tau\to\infty$ limit.
Since $S^*_{\rm WF}$ satisfies $\del_\tau \swf=0$ and the fixed point condition of the WP equation \siki{e:FPCondWP}, all we have to confirm is the vanishing of $\lambda(\infty)$.
As was seen in \Sec{s:FP}, the asymptotic behavior of $\lambda(\tau)$ at the large flow time is controlled by the signature of the quantity $D-2+\eta$.
The anomalous dimension $\eta$ can be explicitly calculated with the $\epsilon$ expansion at this fixed point \cite{Dutta:2020vqo} and is given to $\order{\epsilon^2}$ by
\begin{align}
    \frac{\eta}{2} = \frac{N+2}{(N+8)^2}\frac{\epsilon^2}{4}.
\end{align}
Then, we get 
\begin{align}
    D-2+\eta = 2-\epsilon + \frac{N+2}{(N+8)^2}\epsilon^2
\end{align}
to $\order{\epsilon^2}$ in $D=4-\epsilon$.
Since $\epsilon$ is small within the $\epsilon$ expansion, this quantity is positive.
Recalling that $\lambda(\tau)$ behaves asymptotically as $\tau\to\infty$ like
\begin{align}
    \lambda(\tau) \sim \exp\qty(-\frac{\tau}{2}\qty(D-2+\eta)),
\end{align}
we conclude that $\lambda(\tau)$ vanishes at $\tau=\infty$.
Therefore, we readily find that the action $\swf$ at the WF fixed point satisfies the GFERG equation in the $\tau\to\infty$ limit.

Let us see the relationship between the signature of $D-2+\eta$ and the cluster decomposition principle concretely from the two-point function of the WF fixed point action.
According to \cite{Dutta:2020vqo}, the connected two-point function of $\phi_a$ is given to $\order{\epsilon^2}$ by
\begin{align}
    \ev{\phi_a(p)\phi_b(-p)}^{\rm connected} = \frac{\delta_{ab}}{(p^2)^{\frac{2-\eta}{2}}}
\end{align}
in the momentum space.
By performing the inverse Fourier transformation, we get 
\begin{align}
    \ev{\phi_a(x)\phi_b(0)} \propto  \frac{\delta_{ab}}{x^{D-2+\eta}}
\end{align}
in the position space.
From this equation, we can explicitly confirm that $D-2+\eta>0$ follows from the cluster decomposition principle.

\subsection{Perturbative Solution around WF Fixed Point}
In this subsection, we solve \siki{e:GFERG_NL} to $\order{\epsilon}$ around the WF fixed point and study the flow structure around it.
The solution to the general GFERG equation is already given in \siki{e:SolGFERG}.
In the present case, $D=4-\epsilon$ and $\nm = 3$, and then $S_\tau$ is given to the linear order in $\epsilon,\delta c^A$ and  $\lambda_0^2$ by
\begin{align}
    S_\tau = S^* + \sum_A \qty(\delta c^A e^{x_A\tau'}+\lambda_0^{2}\frac{e^{x_A\tau'}-e^{-(2-\epsilon+\eta)\tau'}}{x_A+2-\epsilon+\eta}h^A)\mc{O}_A.
\end{align}
$h^A$ is defined in the same way as \siki{e:DefhA}, where $H(S^*)$ is given by
\begin{multline}
\label{e:HNL}
H(S^*)  
\coloneqq 
-2e^{-S^*}
\int_x 
\fdv{\phi_a(x)}
\qty(\phi_b(x) + \int_y \mc{D}(x-y)\fdv{\phi_b(y)})
\\ \times 
\del_\mu^2\qty(\phi_b(x) + \int_y \mc{D}(x-y)\fdv{\phi_b(y)})
\qty(\phi_a(x) + \int_y \mc{D}(x-y)\fdv{\phi_a(y)})
e^{S^*}.
\end{multline}

Here we study contributions of some eigenoperators to $S_\tau$ concretely around the WF fixed point.
To this end, we must specify the eigenoperators $\{\mc{O}_A\}$ of the linearized WP equation around it.
The Wilson action is decomposed around the fixed point as $S_\tau=S^*+\delta S(\tau)$, and we use the local potential approximation (LPA), in which the fluctuation $\delta S(\tau)$ takes the following form:
\begin{empheq}[left=\empheqlbrace]{align}
    \label{e:StLPA}
\begin{aligned}
    \delta S(\tau) &=\int_p -\frac{p^2}{2K(p)} \phi_a(p)\phi_a(-p) + \int_x V\\
    V &= \sum_{n = 2}^{N_{\rm max}} \frac{g_{2n}(\tau)}{2^n n!}(\phi_a(x)^2)^n,
\end{aligned}
\end{empheq}
where $g_{2n}(\tau)$ is the $\tau$-dependent coupling of the $2n$-point vertex and $N_{\rm max}$ is the truncation level of the LPA larger than $2$.

By substituting \siki{e:StLPA} into the linearized WP equation with respect to $\delta S$, we can write down the time evolution equation for $g_n(\tau)$ and calculate $\hat{R}$ explicitly.
Then we get a set of its eigenoperators $\mc{O}_A$ by diagonalizing it.
The detailed calculations are shown in \App{a:FlowStructureWF} and we just cite its results here;
$\hat{R}$ has only one relevant operator
\begin{align}
\mc{O}_1=\phi_a(x)^2 +\order{\epsilon} \quad \text{with} \quad x_1 = 2-\epsilon\frac{N+2}{N+8} + \order{\epsilon^2},
\end{align}
and the other local operators are all irrelevant.
An example of the irrelevant operators is
\begin{align}
\mc{O}_2= (\phi_a(x)^2)^2-\frac{4(N+2)}{N}\phi_a(x)^2 +\order{\epsilon} \quad \text{with} \quad x_2=-\epsilon + \order{\epsilon^2}.
\end{align}
Note that this result does not depend on the truncation level $N_{\rm max}$.

Then we can calculate the expansion coefficient $h^A$ for these operators and their contributions to $S_\tau$.
Because the right hand side of \siki{e:HNL} has one factor of the Laplacian $\del_\mu^2$, $H(S^*)$ is expanded with field operators with two or more derivatives.
Thus as far as $\mc{O}_A$ is a linear combination of operators without derivatives like $(\phi_a(x)^2)^n$, its expansion coefficient $h^A$ of $H(S^*)$ is zero.
Therefore, we find that their contributions to $S_\tau$ within the LPA are given by
\begin{multline}
    S_\tau = S^* 
    +\int_x \qty[
    \delta {c}^{1} e^{\qty(2-\epsilon\flatfrac{(N+2)}{(N+8)})\tau'}\mc{O}_1
    +\delta {c}^{2} e^{-\epsilon \tau'} \mc{O}_2
    ]
    \\
    +\sum_{A\neq 1,2} \qty(\delta c^A e^{x_A\tau'}+\lambda_0^{2}\frac{e^{x_A\tau'}-e^{-(D-2+\eta)\tau'}}{x_A+2-\epsilon}h^A)\mc{O}_A.
    \label{e:StNLconc}
\end{multline}

Finally, let us discuss the scaling dimensions of the eigenoperators $\mc{O}_A$ around the fixed point.
As we have seen in the previous paragraph, the expansion coefficient $h^A$ for a linear combination of the field operators without derivatives like $(\phi_a(x)^2)^n$ is zero in the present case of the gradient flow equation \siki{e:GFeqNL}.
Thus, the time dependence of $S_\tau$ in the direction of such operators is just given by $e^{x_A t}$ as seen from \siki{e:StNLconc}.
Because this time dependence agrees with the WP equation, we conclude that such operators have the same scaling dimensions as the WP equation.
For operators with derivatives, which appear when one goes beyond the LPA, their expansion coefficients $h^A$ do not vanish in general.
In such a case, the non-linear terms in the gradient flow equation give a difference between the GFERG equation and the WP equation.
Therefore the scaling dimensions of those operators are different from those of the WP equation.
It should be noted that this result highly depends on the form of the non-linear terms in the gradient flow equation.

\section{Summary and Discussion}
\label{s:summary}
In this paper, we have discussed the GFERG for scalar field theories in general and investigated its fixed point structure.
We have explicitly written down the GFERG equation with an arbitrary gradient flow equation and then discussed its fixed point action.
Remarkably, the fixed points appear for a large flow time limit and are precisely the same as those of the WP equation due to the vanishing of the terms involving $\lambda(\tau)$, originating from the non-linear terms in the gradient flow equation.
Furthermore, we have calculated the scaling dimensions of operators around the fixed points by solving the GFERG equation to the leading order of the deviations from the fixed points and $\lambda_0$.
We find that the relevant operators around the fixed points of the GFERG equation have the same scaling dimensions as those of the WP equation, while the irrelevant operators have different ones generally.
Therefore, critical exponents around the fixed points of the GFERG equation are the same as those of the corresponding fixed points of the WP equation, resulting in the same prediction for its low-energy physics.

We have illustrated these results with a concrete example, the $O(N)$ non-linear sigma model in $4-\epsilon$ dimensions.
In this model, the GFERG equation has the WF fixed point for a large flow time, like the WP equation for the $O(N)$ \textit{linear} sigma model.
Within the leading order of the $\epsilon$ expansion, the Wilson action around the fixed point contains a relevant operator $\phi_a^2$ with the scaling dimension $2-\epsilon(N+2)/(N+8)$ and an irrelevant operator $(\phi_a^2)^2 - (4(N+2)/N)\phi_a^2$ with the scaling dimension $-\epsilon$, which agree with those around the fixed point in the WP equation (up to redefinition of the operators).
These results are consistent with the above general discussion.

There are some open questions to be considered.
As stated above, irrelevant operators in the GFERG equation have different scaling dimensions from those in the WP equation in general.
On the other hand, it is believed that different schemes provide the same scaling dimensions within ERG by a field redefinition.
Thus it seems that \textit{GFERG is not a kind of ERG} but an alternative framework to study the low-energy physics in the Wilsonian sense.
We, however, emphasize that GFERG gives the same prediction on the low-energy renormalized theory as ERG.
This is because they have the same renormalized trajectories and critical exponents around the fixed points.
This point will be confirmed by further studies elsewhere.

As stated above, the Gaussian fixed point can arise in the GFERG equation of the $O(N)$ non-linear sigma model in addition to the WF one.
This fact seems mysterious because the $O(N)$ non-linear sigma model belongs to the universality class characterized by the WF fixed point rather than the Gaussian one.
The existence of the Gaussian fixed point seems extra.
However, we should note that the action at the Gaussian fixed point does not satisfy the constraint $\phi_a^2=\text{const.}$, which is always respected by the GFERG flow.
Thus this Gaussian fixed point is only apparent and that the corresponding flow will never converge to it whatever the initial condition of the GFERG equation is.
In other words, any initial points satisfying the condition $\phi_a^2=1$ at $\tau=0$ do not flow into the Gaussian fixed point for $\tau \to \infty$.

We also comment on our discussion to obtain the fixed points of the GFERG equation in \Sec{s:FP}.
Although the vanishing of $\lambda(\tau)$ is essential there, we have an exceptional case 
in which $\lambda(\tau)$ does not depend on the RG flow time ($d\lambda(\tau)/d\tau=0$), i.e., $\eta_\tau = 2-D$
holds for an arbitrary flow time $\tau$.
In particular, this equation requires $\eta_\tau$ to be zero in two-dimensions.
This means that $Z_\tau$ is also time-independent constant, and we do not need to perform the additional wave function renormalization for the fields $\varphi_i$ to keep their correlation functions UV finite.
An example is the non-linear sigma model in two-dimensions \cite{Makino:2014sta}, with which the GFERG equation associated can have fixed points that are not covered by our argument.

As a possible future direction, it is interesting to consider gauge theories or gravity within GFERG.
Since the most plausible point of GFERG is its manifest gauge invariance,
it would help us to investigate their fixed point structures in a gauge invariant way.
However, the situation there is expected to be different from the case of scalar field theories.
The key point of our analysis is the vanishing of $\lambda(\tau)$, and this quantity should not vanish for gauge theories or gravity for a large flow time.
Indeed, as was discussed in the original paper \cite{Sonoda:2020vut}, the counterpart of $\lambda(\tau)$ in the pure Yang-Mills theory is given by $e^{-\tau(D-4)/2}Z_\tau^{-1/2}$, and becomes $t$-independent constant because $Z_\tau$ can be set to unity in $D=4$.
Thus our present argument is not applied to them straightforwardly and we need more detailed arguments for GFERG in these theories, which is left as future work.

\section*{Acknowledgment}
The authors thank Nobuyoshi Ohta and Katsuta Sakai for useful discussions.
Y.~H. thanks the Theoretical Particle Physics Group at Kyoto University for the kind hospitality during his visit, where part of this work was carried out.
This work is supported by JSPS Grant-in-Aid for Scientific Research KAKENHI Grant No. JP20J11901 (Y.~A.), JP21J01117 (Y.~H.), and JP21J14825 (J.~H.).

%%%%%%%%%%%%%%%%%%%%%
\appendix

\section{Notation}
\label{a:notation}

In this paper, we use the following compact notation for integrals:
\begin{align}
  \int_x \coloneqq \int d^Dx,
  \qquad
  \int_p \coloneqq \int \frac{d^Dp}{(2\pi)^D},
\end{align}
where $p$ denotes a momentum.
The Dirac's delta function in real space is given by
\begin{align}
  \delta^D(x) = \int_p e^{i p \cdot x}.
\end{align}
The Fourier transformation of $\phi(x)$ is
\begin{align}
  \phi(x) = \int_p \phi(p) e^{i p \cdot x},
  \qquad
  \phi(p) = \int_x \phi(x) e^{- i p \cdot x}.
\end{align}
The functional derivative with respect to the field in the momentum space $\phi(p)$ is defined by the Fourier transformation as
\begin{align}
  \fdv{\phi(p)} \coloneqq \int_x e^{i p \cdot x} \frac{\delta}{\delta \phi(x)},
\end{align}
which satisfies the following normalization:
\begin{align}
  \fdv{\phi(q)} \phi(p) = \int_x e^{i (q- p)\cdot x} = (2\pi)^D \delta^D(p-q).
\end{align}
The Laplacian $\del_\mu^2$ is defined as
\begin{align}
  \del_\mu^2 \coloneqq \sum_{\mu=1}^D\pdv{x^\mu}\pdv{x^\mu}.
\end{align}

\section{Comment on Fixed Points of GFERG Equation at Finite Time}
\label{a:finite-time-action}

In this section, we discuss fixed points of the general GFERG equation at the finite flow time.
Because $\del_\tau S^*=0$ is required at an arbitrary time, $S^*$ should satisfy 
\begin{align}
\qty(\phi_a(x) + \int_y \mc{D}(x-y)\fdv{\phi_a(y)})
e^{S^*}=0
\end{align}
in addition to the fixed point condition of the WP equation. 
The solution to this equation can be found easily and given by
\begin{align}
    S^* = -\frac{1}{2}\int_{x,y} \mc{D}(x-y)\phi_a(x)\phi_a(y)
    =-\frac{1}{2}\int_{p} \frac{k(p)}{p^2}\phi_a(p)\phi_a(-p).
\end{align}
The fixed point condition of the WP equation requires $k(p)$ to satisfy
\begin{align}
    \qty(4p^2k(p)+2p^2 \frac{d k(p)}{d p^2} - \eta_\tau k(p))\qty(\qty(\frac{k(p)}{p^2})^2-1)=0
    \label{e:FPCondk}
\end{align}
Because $k(p)$ is given by $K(p)(1-K(p))=e^{-p^2}(1-e^{-p^2})$, it does not satisfy \siki{e:FPCondk} and therefore we have no fixed point at finite time.

\section{RG Flow Structure around Wilson-Fisher Fixed Point}
\label{a:FlowStructureWF}
In this section, we study the RG flow structure of the WP equation around the WF fixed point using the local potential approximation.

Within this approximation, $S_\tau$ takes the following form:
\begin{align}
    S_\tau&=S^*+\delta S(\tau),
\end{align}
where 
\begin{empheq}[left=\empheqlbrace]{align}
    \tag{\ref{e:StLPA}}
\begin{aligned}
    \delta S(\tau) &=\int_p -\frac{p^2}{2K(p)} \phi_a(p)\phi_a(-p) + \int_x V\\
    V &= \sum_{n = 2}^{N_{\rm max}} \frac{g_{2n}(\tau)}{2^n n!}(\phi_a(x)^2)^n.
\end{aligned}
\end{empheq}
Substituting these equations into the linearized WP equation, we get
\begin{align}
\label{e:WPforg}
    \del_\tau g_{2n} 
    = (-2n+4 + (n-1)\epsilon) g_{2n}
    +
    \frac{N+2n}{16\pi^2}g_{2n+2}
    +
    4n m_*^{2} g_{2n}+4n(n-1)\lambda_* g_{2n-2}
\end{align}
to $\order{\epsilon}$.
For simplicity, let us define $\zeta_{n}$ as
\begin{align}
    \chi_{n}(\tau) &\coloneqq \qty(\frac{N}{32\pi^2})^{n-1} g_{2n},
\end{align}
and \siki{e:WPforg} is rewritten in terms of them as
\begin{align}
    \del_\tau \chi_n = \hat{R}_{nm}\chi_m,
\end{align}
where
\begin{align}
    \hat{R}_{nm} &\coloneqq A_{nm}+\epsilon B_{nm},\\
    A_{nm}&\coloneqq 2(2-n)\delta_{n,m} +2\frac{N+2n}{N}\delta_{n+1,m},\\
    B_{nm}&\coloneqq \qty(\frac{2(N+5)}{N+8}n-1)\delta_{n,m}-\frac{N}{N+8}n(n-1)\delta_{n-1,m}
    .
\end{align}

Let us calculate eigenvalues of $\hat{R}_{nm}$ by the perturbation theory with respect to $\epsilon$.
The left- and right-eigenvector of $A_{nm}$ of the eigenvalue $2$ is given by
\begin{align}
\label{e:v1L}
    v_1^L & = \qty(1,0,\ldots,0),\\
    v_1^R & = \qty(1,(v_1^R)_n),
\end{align}
where 
\begin{align}
    (v_1^R)_n = \prod_{m=2}^n\frac{1}{m-1}\frac{N+2m-2}{N}
\end{align}
for $n\geq 2$.
Note that $v_1^R\cdot v_1^L = 1$ and $v_1^L$ corresponds to the operator $\frac{1}{2} \phi_a^2$.
Then, the eigenvalue of $\hat{R}_{nm}$ at $\order{\epsilon}$ is calculated as
\begin{align}
    2 + \epsilon \frac{v_2^R \cdot B\cdot v_2^L}{v_2^R\cdot  v_2^L}
    =2 - \frac{N+2}{N+8}\epsilon.
\end{align}

Next, let us calculate the correction for the eigenvalue $0$.
The left- and right-eigenvector of $A_{nm}$ with the eigenvalue $0$ is given by
\begin{align}
    v_2^L & = \qty(-\frac{N+2}{N},1,0\ldots,0),\\
    v_2^R & = \qty(0,1,(v_2^R)_n),
\end{align}
where 
\begin{align}
    (v_2^R)_n = \prod_{m=3}^n\frac{1}{m-2}\frac{N+2m-2}{N}
\end{align}
for $n\geq 3$.
Note that $v_2^R\cdot v_2^L = 1$ and $v_2^L$ corresponds to the operator $\frac{1}{8} (\phi_a^2)^2 - \frac{N+2}{2N}\phi_a^2$.
Then, the correction to the eigenvalue $0$ at $\order{\epsilon}$ is given by
\begin{align}
    0 + \epsilon \frac{v_2^R\cdot B\cdot v_2^L}{v_2^R\cdot  v_2^L}
    = -\epsilon.
\end{align}

Finally, let us calculate the correction for the eigenvalue $-2$.
The left- and right-eigenvector of $A_{nm}$ with the eigenvalue $-2$ is given by
\begin{align}
    v_3^L & = \qty(\frac{(N+2)(N+4)}{2N^2},-\frac{N+4}{N},1,0\ldots,0)\\
    v_3^R & = \qty(0,0,1, (v_3^R)_n),
\end{align}
where 
\begin{align}
    (v_3^R)_n = \prod_{m=4}^n\frac{1}{m-3}\frac{N+2m-2}{N}
\end{align}
for $n\geq 4$.
Note that $v_3^R\cdot v_3^L = 1$ and $v_3^L$ corresponds to the operator $\frac{(N+2)(N+4)}{4N^2}\phi_a^2 - \frac{N+4}{8N}(\phi_a^2)^2+\frac{1}{48}(\phi_a^2)^3$.
Then, the correction for the eigenvalue $-2$ at $\order{\epsilon}$ is
\begin{align}
-2 + \epsilon \frac{x_2^R\cdot B\cdot x_2^L}{x_2^R\cdot  x_2^L}
= -2 -\frac{N-26}{N+8}\epsilon.
\end{align}
Note that these results hold for any number of the truncation level $N_{\rm max}$.

%% References
\newcommand{\arxivfont}{\rmfamily}
\bibliographystyle{yautphys}
\bibliography{ref}

\end{document}